# Dielectric insulation and high-voltage issues

*D. Tommasini*
CERN, Geneva, Switzerland

**Abstract**
Electrical faults are in most cases dramatic events for magnets, due to the large stored energy which is potentially available to be dissipated at the fault location. After a reminder of the principles of electrostatics in Section 1, the basic mechanisms of conduction and breakdown in dielectrics are summarized in Section 2. Section 3 introduces the types and function of the electrical insulation in magnets, and Section 4 its relevant failure mechanisms. Section 5 deals with ageing and, finally, Section 6 gives some principles for testing. Though the School specifically dealt with warm magnets, for completeness some principles of dielectric insulation for superconducting accelerator magnets are briefly summarized in a dedicated appendix.

## 1 Basic principles

### 1.1 The electric field

This section will summarize the basic principles of electrostatics in a minimalist approach tailored to an introduction to the dielectric insulation of magnets. The reader is invited to check reference textbooks such as, for example, Ref. [1] and the recommended 'further reading'.

We first introduce the *electric field* **E** as a physical entity capable of producing a force **F** on an electrical charge *q*:

$$\vec{F} = q \cdot \vec{E}$$

An electric field can be generated in different ways, in particular by electric charges producing a distribution of *electrical potential V* and/or by a time varying *magnetic field vector potential* **A** according to the Maxwell equation:

$$\vec{E} = -\nabla V - \frac{\partial \vec{A}}{\partial t}.$$

As the Earth's surface is electrically equipotential, it is convenient to set its electrical potential to zero (electrical ground).

It is important to remark, and this is the main basis of a dielectric insulation system, that what produces forces on electrical charges, responsible for creating electrical currents, is the electric field and not the absolute value of the electric potential. A bird can have a rest on a high-voltage transmission line because the electric field is not high enough to trigger an electrical discharge.

### 1.2 Definition of 'dielectric'

The word *dielectric* comes from the Greek 'dia = through' + 'electric', which was condensed into 'dielectric' for ease of pronunciation. In 1836 Faraday [2] discovered that electric charges created by a high-voltage generator could not create an electric field inside a room enclosed by a metallic envelope (what is since called a Faraday cage). In practice, electric field lines do not 'pass through' an electrical conductor, in opposition to what happens with any material not carrying electricity (like glass or air). Faraday thus needed a new term to define such 'non-electrical-conducting' materials allowing the electric field to pass through and consulted William Whewell who, in December 1836, invented the term *dielectric*.

It is interesting to note that Whewell [3] coined an impressive series of other important scientific terms, in certain cases by analogy to other words (*scientist*, *physicist* in analogy to the word *artist*), in other cases to help friends facing specific scientific issues (among others Lyell with the terms *Eocine*, *Miocene* and *Piocene* and Faraday with the terms *anode*, *cathode*, *dielectric*, *diamagnetic*, *paramagnetic*, *ion*, *electrode*).

## 1.3    A basic reminder of dielectric properties of matter

In a dielectric material the conduction and valence bands are separated by a large energy gap so that there are no electrons available for electrical conduction: the interaction between an external electric field and the charges present in the dielectric is described by the so-called *polarization*. This phenomenon is the basis of the explanation Maxwell gave to the ability of a capacitor of storing electrical energy and then providing this energy in the form of electrical current. According to Maxwell, the electric field present in a capacitor deforms the distribution of charges (in some way stretches the atoms/molecules) of the material between the electrodes. When the electric field changes in time, the distribution of charges changes, resulting in what he called a *displacement current*. The total current density flowing through a material can then be expressed as the sum of a conduction and a displacement current density according to the formula

$$J_{tot} == \sigma \cdot E + \frac{\partial D}{\partial t},$$

where σ is the electrical conductivity and $D$ is the total polarization effect or *electric displacement field*, which can be expressed as

$$D = \varepsilon \cdot E,$$

where ε is the dielectric permittivity of the material.

The dielectric permittivity describes the ability of a material to polarize when submitted to an electrical field, and is typically referred to that of vacuum according to the relationship

$$\varepsilon = \varepsilon_0 \cdot \varepsilon_r,$$

where $\varepsilon_0 = 8.854 \times 10^{-12}$ F/m is the 'dielectric constant of free space' and $\varepsilon_r$ the relative dielectric constant of the material.

The interested reader is invited to revisit standard textbooks on the theory of dielectrics. We conclude here by recalling the main modes of polarization.

- Electronic polarization: the electric field modifies the electron density. This mode is important in certain crystals such as Si and in round noble gases. The response to the applied field is very fast.

- Orientation polarization: it is important, for example, in water in which the molecules form electric dipoles randomly oriented. The orientation of these dipoles by an electric field is disturbed by thermal noise viscosity: the response to *E* is then slower than the previous one and more dissipative.

- Ionic polarization: it mainly includes bulk and interfacial effects. It is slow and dissipative, and typical of ionic crystals (such as NaCl) and heterogenous systems.

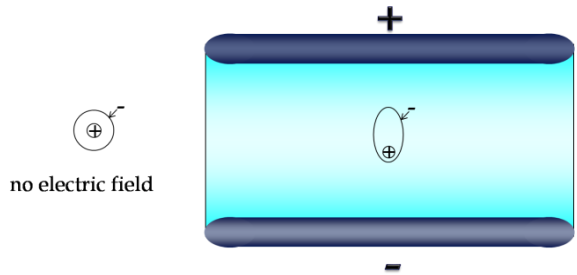

**Fig. 1:** Stretched electron density relevant to electronic polarization

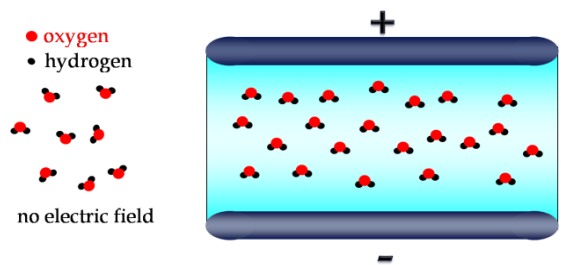

**Fig. 2:** Orientation polarization of water molecules. Electrons in the covalent bond tend to get slightly denser close to the oxygen atom, which is more electronegative than the hydrogen ones

All these modes are dissipative: changing a charge distribution in a dielectric dissipates energy, resulting in heating the dielectric. In the theory of dielectrics these effects are described as dielectric losses and depend on several parameters, in particular on frequency.

Let us take the example of the orientation polarization of water. If the magnet field changes extremely rapidly in time, the molecules are not capable of following the orientation, and the dielectric losses associated with the field change will be small. Very slow changes in time will also produce small losses because the dissipative cycle is repeated only a few times in the unit of time. In the case of water, at ordinary temperatures the maximum of the dielectric losses are at about 10 GHz, which would suggest using this frequency for microwave cooking. In reality, for this application, the frequency is lowered to 2.45 GHz in order to avoid heating only the first layers of water, allowing time to uniformly heat the food.

### 1.4 Electrostatics for high-voltage engineering

A treatment of the theory of electrostatics is beyond the scope of this lecture. We recall the essential concepts to underline the importance of geometries and interfaces in dielectric insulating systems.

As we have understood from the above sections, what counts to exert forces on electrical charges is the electric field, capable of producing conduction and displacement currents. From the Maxwell equation in Section 1.1 the electric field distribution $E$ is computed from the electric potential $V$ and the magnetic potential $A$. Considering the static case, we just need the distribution of the electric potential $V$, which is computed from the *Poisson* equation

$$\nabla^2 V = -\frac{\rho}{\varepsilon},$$

where ρ is the electrical free charge density. In the absence of free charges the *Laplace* equation holds:

$$\nabla^2 V = 0.$$

Let us consider a 'plane capacitor' configuration, with infinite parallel electrodes separated by a distance $d$, at potentials $V_1$ and $V_2$, similar to the one in Fig. 3.

Owing to the plane symmetry, the Laplace equation is mono-dimensional and can be solved as

$$\frac{\partial^2 y}{\partial x^2} = 0 \rightarrow V(x) = c_1 x + c_2 \rightarrow c_2 = V_1; c_1 = \frac{V_2 - V_1}{d}$$

and the electric field is

$$E = \frac{\partial V}{\partial x} = \frac{V_2 - V_1}{d}.$$

This means that, in case of parallel large electrodes the electric field is constant and uniformly distributed between the electrodes. In real cases, as will be shown with a few examples, the electric field will be higher than that obtained with this formula, for different reasons:

- the shape of the electrode is often far from an ideal case of parallel large electrodes (Fig. 3);
- the material between the electrodes is not homogeneous (Fig. 4).

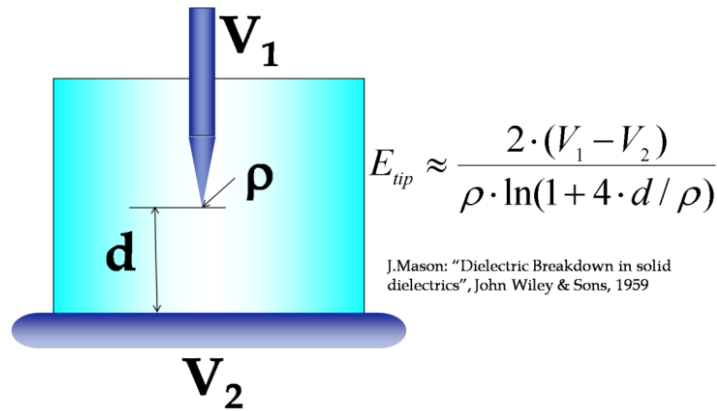

$$E_{tip} \approx \frac{2 \cdot (V_1 - V_2)}{\rho \cdot \ln(1 + 4 \cdot d / \rho)}$$

J. Mason: "Dielectric Breakdown in solid dielectrics", John Wiley & Sons, 1959

**Fig. 3:** The electric field of a 'needle-to-plane' electrodes configuration at the needle tip

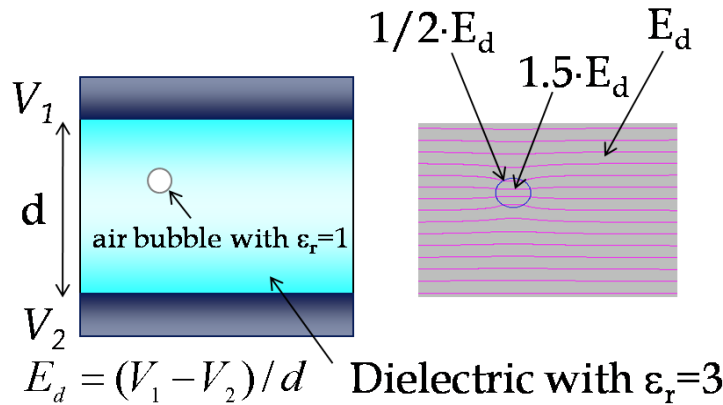

$$E_d = (V_1 - V_2)/d$$

**Fig. 4:** The electric field is enhanced where the dielectric constant is lower, as in air bubbles possibly present in a dielectric insulator

## 2  Breakdown in dielectrics

We have seen in Sections 1.2 and 1.3 that a dielectric is a *non-electrical conducting* material, in which macroscopic electrical currents are mainly due to the displacement current.

However, when the electric field is strong enough, the dielectric material may suddenly lose its property of non-conduction, permanently or temporarily, showing an *electrical breakdown.*

In practice we can define the electrical breakdown as an abrupt rise of electrical current under the effect of an electric field. Its causes depend on the medium, the environmental conditions, the geometry, type and material of the electrodes, and on the type and amplitude of the electric field.

The maximum electric field achievable in a dielectric without the occurrence of an electrical breakdown is called *dielectric strength*, typically expressed in kV/mm.

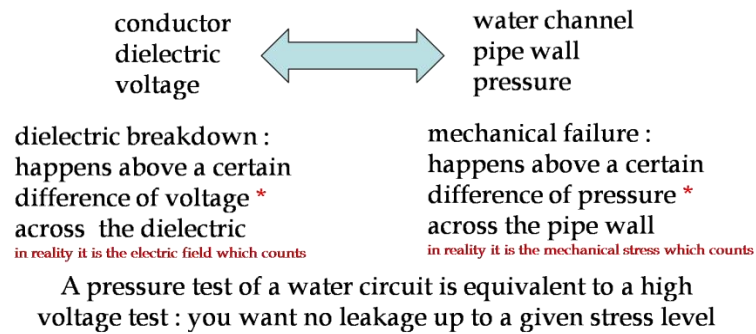

**Fig. 5:** A non-rigorous but effective analogy between dielectric insulation and water hose wall

### 2.1  Breakdown in gas

In a gas, free electrical charges under a sufficiently high force can produce ionization and avalanche breakdown by hitting other atoms. At a given temperature, the relationship between the voltage difference $V_B$ between the electrodes at which a breakdown occurs, their inter-distance $d$ and the pressure $p$ of the gas follows the so-called *Paschen law* [4]:

$$V_B = \frac{a \cdot (p \cdot d)}{\ln(p \cdot d) + b}$$

where the constants $a$ and $b$ depend mainly on the gas.

For each gas, at a given temperature, the curve described by this law has a minimum voltage difference between the electrodes below which no breakdown can occur whatever the product between inter-electrode distance and gas pressure. On the other hand, above that voltage difference, one is likely to find combinations of inter-electrode distance and gas pressure at which an electrical breakdown can happen. This is why using gas as a dielectric, though in general very efficient at high pressures or at very low pressures (high vacuum), can become unsafe if during the operation the gas pressure can vary over a large range. Figure 6 shows the so-called Paschen curve for dry air at 20°C, extrapolated from several experiments (from T.W. Dakin, Electra N°32, 1974). At high pressures a further increase of pressure increases the density of the gas but also decreases the mean free path: though the probability of collisions increases, the lower collision energy due to the shorter mean free path is dominant and provides an increase of $V_B$. At low pressures a further decrease of pressure increases the mean free path but also decreases the density of the gas: though the collision energy increases due to the longer mean free path, the lower probability of ionizing collisions provides an increase of $V_B$. In summary, at high pressures the mean free path is dominant, at low pressures the probability of collision is dominant, and there is a product $p \cdot d$ at which $V_B$ has a minimum.

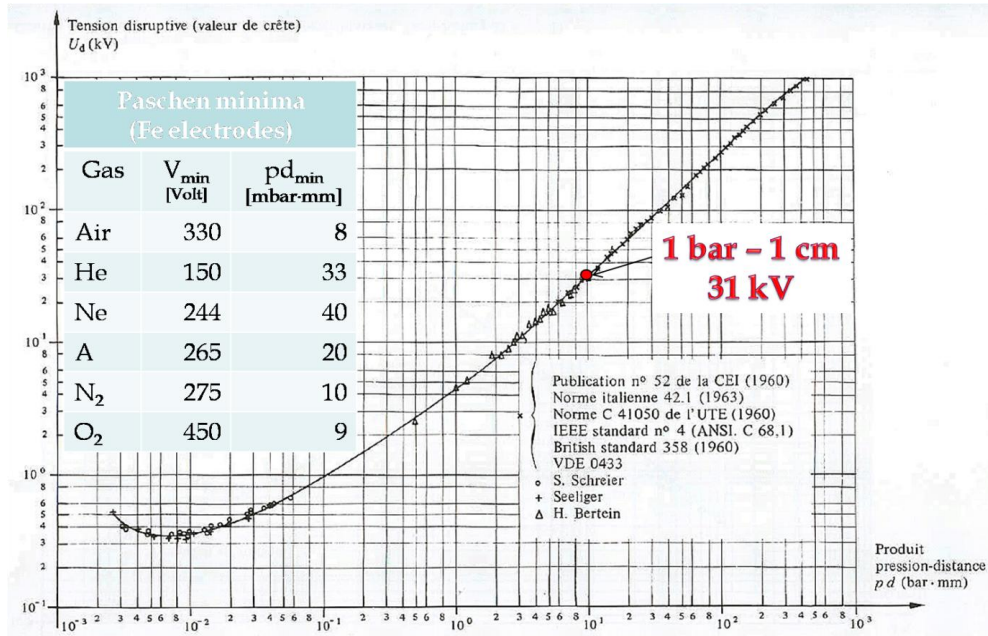

**Fig. 6:** Paschen curve for dry air at 20°C and Paschen minima for different dry gases

## 2.2  Breakdown in liquids

Electrical insulating liquids, in general oils, are used in high-voltage transformers, capacitors, switches, and circuit breakers. In many cases they act as both dielectric insulation and coolant. In accelerator components they are used, for example, in fast pulsers for voltages above 30–50 kV. The most important aspect to consider when dealing with liquid insulation is its purity. When an electric field is applied to an electrical insulating liquid, current is initially dominated by impurities. At fields higher than 100 kV/cm electron emission can start at the interfaces of the impurities triggering an ionization process possibly leading to an electrical breakdown.

The breakdown mechanisms in liquids are not yet fully understood. Experimentally the relationship between the voltage difference $V_B$ between the electrodes and their inter-distance $d$ takes the form

$$V_B = A \cdot d^n$$

where $n<1$ (increasing the distance between the electrodes increases the breakdown voltage less than linearly: this can be justified with statistical considerations on discharge path), and $A$ is a constant.

## 2.3  Breakdown in vacuum

Does an absence of particles mean no breakdown? In principle yes, but practical vacuum still has particles, or residual gases: in these terms breakdown in vacuum is studied together with the breakdown of gases (remember the Paschen law). In practice, only pressures lower than $10^{-2}$ mbar can be considered to provide a real dielectric insulation. At higher pressures, especially in the mbar range, such vacuum can even create ideal conditions for electrical discharge. Particular care must be taken to avoid 'higher pressure' spots for example due to local degassing. In the worst cases, conditions for sputtering can be created, possibly with the help of the magnetic field produced by a magnet, depositing metallic coating on the surface of insulators leading to the catastrophic failure of the insulation. In practice, at low pressures, many parameters become extremely important, such as the pressure itself, the distance between electrodes, the applied voltage, the electrode materials, the type and cleanness of the insulating materials.

## 2.4 Breakdown in solids

The father of the theories of dielectric breakdown in solids is Herbert Fröhlich, a German physicist who, due to the anti-Semitic laws, performed most of his work on dielectrics at the University of Bristol in England. Fröhlich was a theoretical physicist who decided to engage himself in the physical understanding of phenomena like the dielectric breakdown which appeared at that time governed by empirical facts. The review 'Dielectric breakdown in solids' of 1939 [5] still remains a milestone for both scientific merit and clarity. We can't help providing an extraction of his brilliant introduction below:

> *Owing to its great technical importance, the dielectric breakdown in solids has for many years been a subject of experimental and theoretical investigations. Nevertheless, only in recent years has it been possible to come to a closer understanding of this phenomenon. It is the aim of this article to give an account of these recent developments. One of the most important results of recent research has been the experimental proof of the existence of an intrinsic electric strength. This means that at a given temperature a maximum breakdown strength exists for each dielectric substance which is a constant of this substance, and which is obtained under ideal conditions (homogeneous field, uniform material without weak spots. etc.). Therefore it should be, and has been, possible to calculate this intrinsic electric strength from simple physical constants of the material. In this Report we shall deal mainly with the intrinsic electric strength. There exists also a quite different type of breakdown, the so-called breakdown through thermal instability. This type of breakdown is of importance only in special conditions such as high temperature. We shall give only a very short account of the thermal breakdown, since several reports on it have been published already. In addition, there are the various complex forms of breakdown which occur in industrial insulation; but these, although of great engineering importance, usually reduce finally to one of the two fundamental types considered. The description of industrial breakdowns would be lengthy and is related to the particular properties of complicated substances. Since, also, there is an extensive engineering literature on the subject, it will not be treated here.*

The two fundamental types of breakdown mentioned by Fröhlich are:

- the electronic breakdown. With a sufficient energy (above a critical electric field) electrons can cross the forbidden gap from the valency to the conduction band, eventually producing collisions with other electrons and leading to breakdown;
- the avalanche breakdown. As in gases, with sufficient energy (above a critical field) conduction electrons gain enough energy to liberate electrons from the lattice atoms by collisions.

In both cases the breakdown permanently modifies the matter of the failing path.

In practice the electrical breakdown appears below, sometimes much below, the intrinsic limit of the material. Parameters and phenomena affecting the dielectric strength are:

- material type and characteristics (of course)
- type and duration of the applied electric field

- mechanical integrity: in most cases it is a mechanical failure, like a crack, which triggers a fault
- radiation damage: leading to a mechanical failure or to a change of dielectric properties
- chemical actions (like oxidation): typically triggered or accelerated by radiation, temperature
- hydrolysis: in particular when the insulating system is not water-tight
- contamination
- interfaces: in many cases they are a favourite path for a breakdown. At an interface the electric field can be increased due to variations of the dielectric constant, for example, in case of presence of air bubbles, the dielectric strength can be reduced, for example, due to presence of humidity or free ions, as a mechanical discontinuity and interface may lead to a mechanical failure
- progressive erosion of the material creating a breakdown path: starting from air bubbles or propagating along surfaces/interfaces, having different forms (treeing, tracking, partial discharges)
- ageing: some of the above phenomena/parameters produce a progressive transformation over time

## 3   Electrical insulation in magnets

In an electrical machine, the electrical insulation ensures that current flows only along the conductors and not between individual conductors or between the conductors and the other part of the magnet. We distinguish electrical insulation:

- between coil turns
- between different active parts
- between active parts and ground

A weak electrical insulation may produce:

- current leak with local heating up to melting and possible fire
- progressive damage of the leakage path up to a short circuit
- unbalance of circulating current (possibly with magnetic field distortion)
- autotransformer effect with reduction of magnetic field
- incorrect functioning of protections

We shall keep in mind that the energy stored in magnet circuits is available for any catastrophe!

Solid dielectric materials can be distinguished in three main classes:

- inorganic materials: ceramics, glass, quartz, cements and minerals such as mica, etc.
- organic materials: thermoplastic and thermosetting
- composites: fully organic (aramidic fibres-epoxy tapes) or mixed (epoxy-mica tapes)

We will not go through the hundreds of materials possibly suitable for an electrical insulation: we invite the reader to check textbooks and catalogues and remain open to search for the best solution. We are insisting on this because the knowledge and use of the appropriate material and manufacturing technology is one of the most difficult issues in the science and technology of dielectric insulation.

Too often an easy way is chosen, copying the most popular solutions even in cases for which a different specific material would be more appropriate. Unfortunately, it is a matter of fact that the experience gained in the technologies used for large electrical machines and in other high-voltage devices is not always exploited in magnet manufacture. An example, to make just one, is electrical tracking, which will be treated later, requiring the use of specific material in case the risk of contamination/humidity is important. The same magnet, depending on its operation mode, environmental conditions including radiation and humidity, type and amplitude of the supply voltage, mechanical stresses, may require different materials and technologies for its electrical insulation. It is also important to remember that there are no 'definitive' solutions. For example, cables in nuclear plants or accelerator magnets close to target areas are submitted to very high radiation doses and are typically insulated with MgO powders. Their insulation, being mineral, is extremely resistant to radiations, however, it is also very hygroscopic and, though it is typically sealed, it is in general not appropriate for voltages above 1000 V.

## 4     Failure mechanisms

A dielectric insulation is stressed by several factors, among others: electric (field strength and type), thermal, mechanical, chemical (including oxidation), radiation, and of course contamination.

These factors can produce short and/or long term degradation.

Also, and this is unfortunately in many cases not considered with enough attention, environmental conditions can modify the dielectric system leading to its failure not really because of degradation or ageing, but because in operation the properties of the dielectric system are different from the ones the designer has considered. A typical example is the dielectric insulation in proximity to the hydraulic connection posts, which should in principle be capable of providing the required dielectric strength even in presence of humidity.

That said, in magnets the causes of dielectric failures (electrical short circuit) can be basically schematized in three main groups:

- *insufficient dielectric properties (wrong design)*
- *mechanical failures* breaking the dielectric integrity
- *modification of the bulk or surface dielectric properties*

In case the above events are due to an irreversible, progressive modification of the properties of the dielectric system due to external stresses we speak of *ageing*.

It is not easy to universally define the concept of ageing with respect to nominal characteristics of a material. In this sense, we will give here two examples of different nature:

- the difficulty in defining an 'intrinsic' dielectric strength of materials. When an electrical field is applied, if this is above a certain value, even within very short times electric stress can progressively modify the bulk of a dielectric leading to an electrical breakdown. This is why the 'intrinsic' dielectric strength, which should be the one the material can withstand before being modified, is sometimes measured with very short pulses, even just a few nanoseconds, obtaining for certain materials values up to several MV/mm. However, shall we really speak of ageing when the degradation takes place over such a short time? Wouldn't it be better to define an operating dielectric strength, and ageing be defined with respect to that value? This is the practical approach used by most International Standards, and ageing in practice becomes the result of a process taking a reasonably long time and a macroscopically reasonable progressive modification of properties.

- Let us now consider a short circuit that is due to the deposition, along the surface of a dielectric, of humidity. The overall dielectric properties of the insulating system change with respect to a dry situation, however, we believe this is certainly not ageing at least because the modification of the property is not irreversible and, most of all, it is not a modification of the dielectric itself but it is a modification of the insulation system. In case humidity would progressively penetrate in the dielectric the opinions may be more easily shared between ageing and a modification of the system due to an additional element (the water).

A comprehensive overview of failure mechanisms, degradation, and ageing is far beyond the scope of this note and can be found in specialized literature.

We prefer here to briefly recall in the next paragraphs some remarkable examples of failure mechanisms relevant to magnets insulation, which in certain cases are progressive and also fall into the class of ageing. In the next section we recall the basics of the theory of ageing of dielectric materials and insulation systems.

## 4.1 Mechanical failure

A mechanical failure, even partial or very localized like a hole or a crack, is often the reason for an electrical short circuit. Mechanical failures may be produced in many different ways, from the embrittlement of the material due to radiations to mechanical fatigue. Here we want to stress the importance of the quality of the manufacture by discussing a photo (Fig. 7) taken by the author on a sample of resin–fibreglass composite after a flexural fatigue test.

In the picture it is clearly visible how, on that specific specimen, the adhesion between glass fibres and resin was not good, leading under a mechanical flexural strain to local debonding and consequent concentration of stresses, responsible for premature breakage of certain lots of samples. In a dielectric system, for example, a coil insulation on magnets, the quality and treatment of the fibres to provide a good adhesion to the resin, the type and quality of the resin, the curing cycle, and all procedures to avoid contaminants, discontinuities, critical interfaces, air bubbles, are of extreme importance. Depending on the application, the environmental conditions, the required viscosity during the manufacture, the geometries, different preparation of a similar resin compound may be envisaged, acting for example on the base resin (aromatic, cycloaliphatic, novolak or phenolic), the hardener (amine, anhydride), the accelerator, the flexibilizer, the fillers ($Al_2O_3$, $MgO$, quartz, dolomite) and plenty of other additives that modern technology makes available to us.

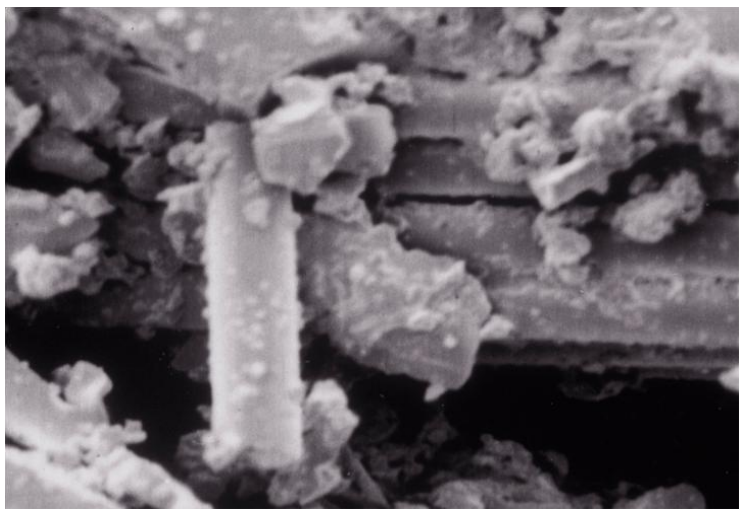

**Fig. 7:** Electron microscopy of a resin–fibreglass composite after a flexural fatigue test (from: G. Liberti and D. Tommasini, 'Fatigue on fibreglass/resin composites', ICCM, Milan, May 1988)

## 4.2 Partial discharges

Partial discharges (PD) constitute one of the main mechanisms of ageing, leading to failure of an electrical insulation in continuous operation. In Fig. 4 we already illustrated how interfaces between air and dielectric such as air bubbles and delamination represent volumes where the electric field is higher with respect to that of the surrounding dielectric. This is due to the difference of dielectric permittivity of the insulating material and that of air. But, in these air regions, in addition to the electric field being higher, the dielectric strength of air is much lower than that of the dielectric. If in the air bubble the electric field exceeds the dielectric strength of air, which is about 31 kV/cm for dry air, and lower in presence of contaminants or humidity, then an electrical discharge appears in the bubble, called *partial discharge* because it does not link the two electrodes. In principle a partial discharge would not cause a problem, because it is not a short circuit. For example in aerial transmission lines partial discharges, in the form of the so-called *corona* effect, can be present permanently during operation, producing some dissipation of energy but no or minor degradation of the system. But this is in free air, where the molecules interested by the PD are continuously 'refreshed' by new ones without the risk of accumulating discharging elements possibly crossing the two electrodes and producing a real short circuit. For information, we remark that corona can be visible in two different colours:

- a positive voltage creates a uniformly distributed bluish-white cloud,
- a negative voltage creates reddish, pulsing spots.

In a solid dielectric the situation may become critical. The PD in the bubble produce ionization capable of breaking the chemical bonds of the dielectric at the boundary of the bubble, will heat the dielectric, will produce carbonization, will increase the air pressure in the bubble, may trigger chemical degradation: in summary PD will progressively erode the internal surface of the bubble either uniformly or by forming channels until (Fig. 8) a complete electrical short circuit is produced.

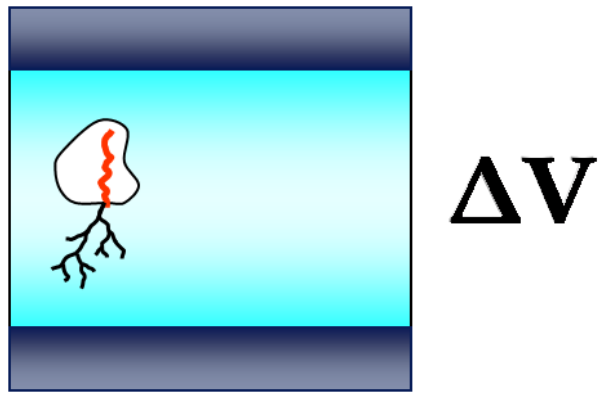

**Fig. 8:** Schematic representation of the propagation of partial discharges in an air bubble

## 4.3 Treeing

Treeing constitutes a progressive propagation of a discharge path produced by partial discharges in case of diverging electrical fields, as for example schematically represented in Fig. 3, where the electric field at the needle tip can be orders of magnitude higher than the one corresponding to the plane electrodes geometry at the same inter-electrode distance and the same voltage difference. Treeing typically affects high-voltage cables, in particular in coaxial geometries, where the electric field is higher close to the central conductor than at the outer cable side. If, in proximity to the central conductor, the electric fields exceeds the dielectric strength of the material (in particular if air is present at the interface between the central conductor and the dielectric, triggering partial discharges), then a branched conductive path will progressively extend from the inner electrode to the outer one. It often happens that high-voltage cables, after several years of operation, show visible electrical treeing which did not yet produce a short circuit.

### 4.4 Tracking

The progressive extension of a conductive path from one electrode to the other with a spread of sparking channels may happen along a surface: in this case we speak of *tracking* (Fig. 9). This phenomenon is typically triggered by a contamination of the surface of the dielectric, which can be for example humidity, creating quasi-equipotential regions between the electrodes in case the contamination is even slightly electrically conducting, increasing then locally the electric field, possibly further enhanced by the difference of dielectric permittivity between the dielectric and air. A proper choice of the dielectric material depending on the operating conditions is crucial every time tracking may possibly become an issue, as for example in magnet connection posts (Fig. 10).

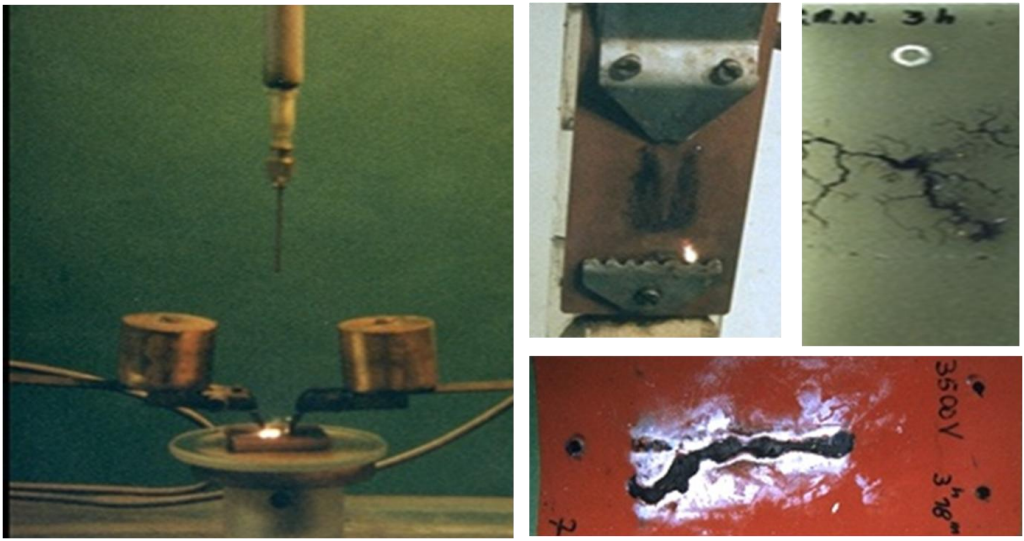

**Fig. 9:** Examples of tracking test set-up and tracking patterns

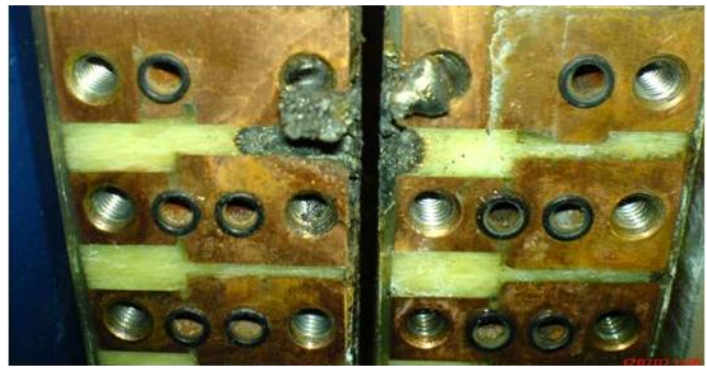

**Fig. 10:** A failure at a magnet connection post due to tracking triggered by humidity

### 4.5 Radiation

We can distinguish radiation from:
- charged and neutral light particles
- charged and neutral heavy particles
- electromagnetic radiation

These can excite electrons and, in case they have enough energy (typically more than 10 eV) they may produce ionization. The ICNRP (International Commission of Non-Ionizing Radiation Protection) defines electromagnetic radiations, from radiofrequency to ultraviolet, as well as electric and magnetic fields, as *non-ionizing radiations*. In particle accelerators we are typically confronted with *ionizing* radiations, for which we recall two basic concepts: the *range*, consisting in the distance over which a particle of given type and energy loses most of its energy, and the *stopping power* which represents the energy loss per unit of distance covered by the particle. While for electromagnetic radiation and for light particles the stopping power decreases over the path in matter, for hadrons it has a peak, called *Bragg peak*, towards the end of the range. This property is being used in recent medical accelerators to kill cancerous cells (hadron therapy).

Concerning the effects in dielectrics, the electrons produced by the ionization can excite molecules and break bonds forming free radicals, very reactive especially when oxygen is present. New cross-links can also be formed, possibly leading to a modification of the structural properties of the material, which in general becomes stiffer but also more brittle. The covalent bond, typical of polymeric materials, is very sensitive to ionizing radiations.

As in many cases the effect of ionizing radiations depends mostly on the total radiation energy effectively transmitted to the material, we define the *absorbed* dose, expressed in *Gray*, as the energy effectively absorbed by a material per unit of weight:

$$1 \text{ Gy} = 1 \text{ J/kg} (= 100 \text{ rad}) .$$

In general, relatively to radiation, failure of a dielectric insulation is due to loss of mechanical properties (in particular embrittlement) or to the evolution of gases inside the material.

The collection of tables in Fig. 11, extracted from several CERN publications, reports a classification of the resistance of materials with respect to radiations. The tables are barely readable, their intent is to invite the reader to check directly these publications, which provide details on the measurement methods and the criteria used to compile this classification.

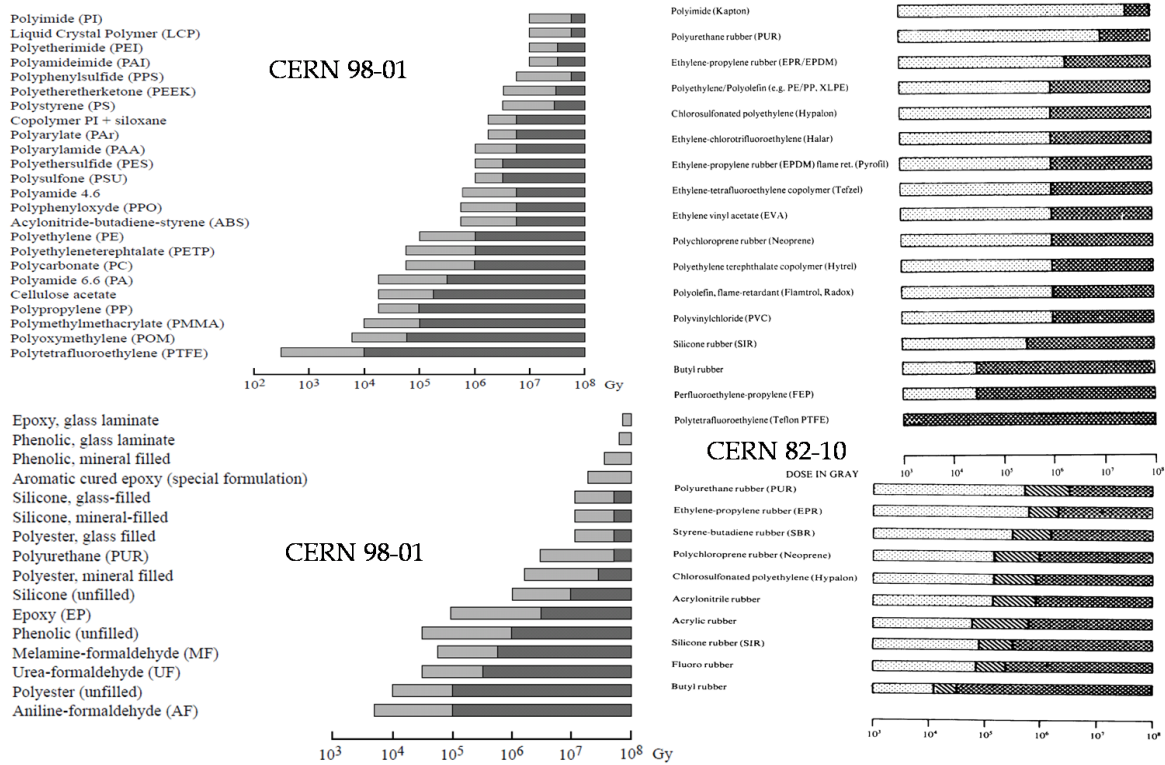

**Fig. 11:** Compilation of radiation damage data in accelerator materials and components

However, we also know that in certain cases, especially for heavy ions, the electrical properties can be heavily affected by ionizing radiation because of formation of free charges or change of energetic levels of the matter.

Figure 12 refers to the results of an experiment recently carried out in the frame of a common scientific activity between CERN and GSI to understand the difference of electrical damage produced, at given absorbed doses, by different ionizing radiation on polyimide tapes. The effect of heavy-ion radiation on the dielectric strength is noticeable already at very low radiation doses.

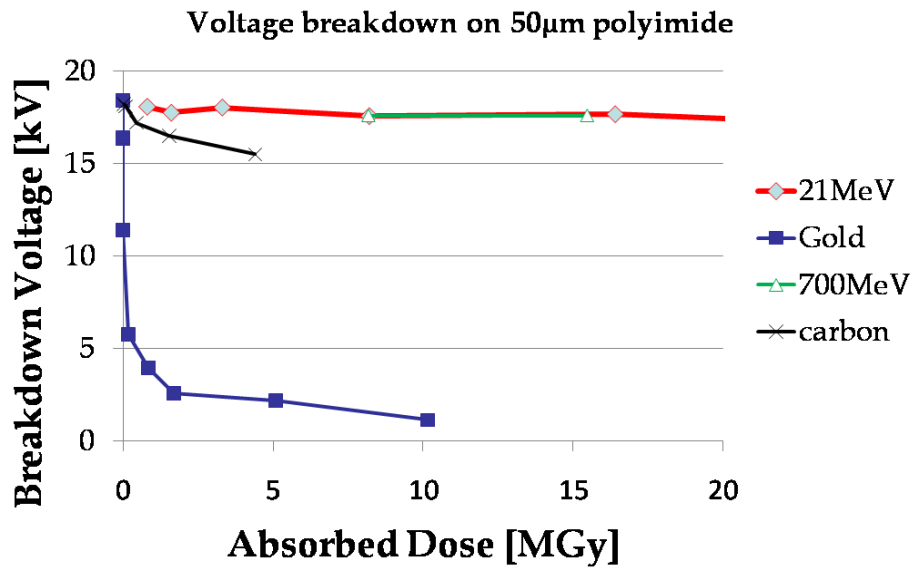

**Fig. 12:** Dielectric strength of irradiated polyimide [courtesy R. Lopez, CERN, and T. Seidl, GSI]

To conclude this short introduction to radiation damage of dielectrics, we recall situations where radiation doses are extremely high [above 10–100 MGy]. In these cases the technologies used in the magnets have to be robust towards the effects of such high radiation levels, and also the techniques to be used for a possible replacement of a defective magnet have to be conceived at an early stage of the magnet design to reduce the exposure of people to radiations.

Such high levels of radiation require in general the use of special materials:

- inorganic insulations (issues: bonding and moisture absorption)
    - cements and minerals (concrete, mica, quartz)
    - glasses
    - ceramics (oxides of aluminum, magnesium, beryllium)
- new compounds (issues: mechanical properties still to be fully confirmed over a long time)
    - cyanathe esther
    - blends epoxy/cyanathe esther

We recall here the CERN report 82-05 (Fig. 13) for mineral materials and recent developments being carried out in the frame of the ITER project for new compounds.

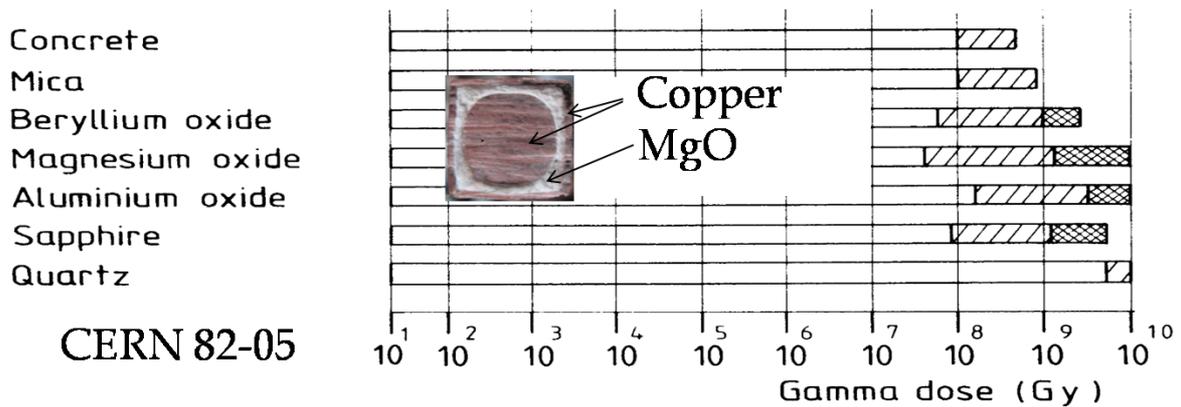

**Fig. 13:** Radiation damage data for mineral insulations; cross section of a MgO insulated cable.

## 5 Ageing

We already discussed the definition of ageing, and we have also introduced in the previous sections some of the processes producing ageing. In this section we will recall the basic principles of the theory of ageing, taking as example the thermal ageing, possibly triggering a fault as a consequence of the modification of dielectric properties, and the electrical ageing.

The purpose of this approach is to introduce the reasoning principles applicable to the design and evaluation of dielectric systems for magnets.

### 5.1 Thermal ageing

The modification of the properties of the material is due in this case to the effects of temperature over time. The effects of thermal ageing were identified and empirically quantified by Montsinger [6], who found that the degradation speed of paper oil insulated transformers doubles every 8°C of temperature increase above the operating temperature of the device.

This finding is generally known as the *Montsinger's rule*, stating that:

*a temperature rise of 10°C halves the expected lifetime of an electrical system*

Montsinger was also convinced that the mechanism of thermal ageing led to a mechanical failure of the dielectric, which at a certain moment develops a crack representing a path for an electrical discharge. Furthermore, he was not convinced of the idea expressed by Steinmez and Lamme in 1913 [7] that below a certain temperature threshold there is no ageing: according to his idea, as the lifetime halves every temperature increase of 8–10°C, it should also double at every temperature decrease of the same amount.

The real basis of thermal ageing were assessed by Dakin in 1948 [8], who proposed that thermal ageing is due to chemical reactions, like oxidation, modifying the chemical structure of the material. In that case, their speed shall be governed by the Arrhenius law:

$$V_R = A_R \cdot e^{-\frac{E}{KT}}$$

where $V_R$ is the velocity of the chemical process, $E$ is its activation energy, $T$ the absolute temperature, $K$ the Boltzman constant, and $A_R$ a constant.

Starting from this assumption, Dakin could show that the *lifetime L* of an electrical system submitted to thermal ageing, defined as the time of operation at a given temperature at which a given property (for example the dielectric strength) goes below a given threshold can be expressed in the simple form:

$$L = C \cdot e^{\frac{B}{T}}$$

where *C* and *B* are constant.

This allows one to plot the so-called 'lifetime curves' as a function of temperature which, in an appropriate scale, are straight lines: 'Arrhenius lines or Arrhenius plot':

$$\ln L = \ln C + \frac{B}{T}.$$

This formalization opened the way to accelerated tests, because, in principle, once the two constants *C* and *B* are determined by measuring the lifetime at two temperatures (higher than the operating one, to complete the test in a short time), it is possible to obtain the lifetime at any other temperature (Fig. 14).

Dakin understood immediately the potential of his model, but also correctly warned that this should be used within a reasonable range of temperatures, in order that the process remain the same.

For example, below temperatures not capable of activating the chemical process responsible for ageing there is no modification of the material properties, solving the contradiction of Montsinger's approach towards the absence of thresholds. On the other hand, at very high temperatures, instead of for example a progressive oxidation we may directly burn the material.

In practice Dakin's model is used to define the so-called *temperature classes*. The temperature class of an electrical material or system is in general defined as the temperature below which the lifetime exceeds 20 000 hours (IEC 60216-1).

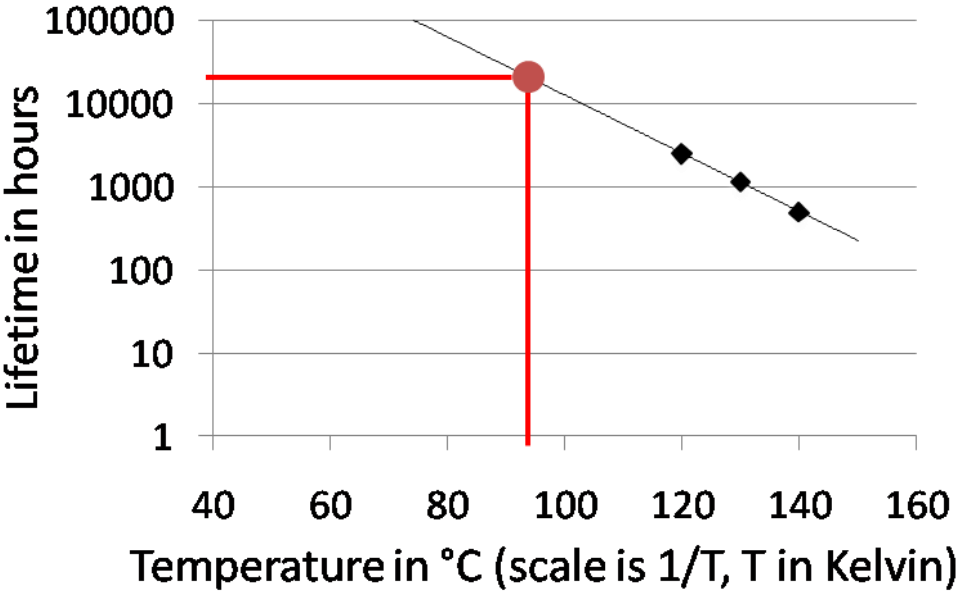

**Fig. 14:** Arrhenius plot and its use to extrapolate lifetime. The three black squares are experimental lifetime measurements after ageing at three different temperatures, the red lines refer to the extrapolation at a lifetime of 20 000 hours. In this example the temperature class is 90°C.

## 5.2 Electrical ageing

In this case the modification of the properties of the material is due to the effect of the electric field. Electrical ageing is in general triggered by defects, for example treeing in high-voltage cables, in particular in ac where the alternating electric field may also induce dielectric heating.

The study of 'pure electrical' ageing is complicated and requires very high electric fields, typically produced by divergent fields as 'needle-plane' geometries, in practice falling into electrical treeing.

In general the lifetime *L* of a material undergoing electrical ageing can be empirically described by the relationship:

$$\mathbf{L} = C \cdot (E - E_0)^{-m}$$

where *C* and *m* are constant, *E* is the applied electric field and $E_0$ a threshold electric field.

This relationship recalls that of mechanical fatigue, where the lifetime can be replaced by the number of cycles and the electric field by the mechanical stress, such that we may speak of 'electrical fatigue'.

## 6 Testing

An electrical insulation shall be tested in such a way as to be confident that it is capable of providing the required insulation levels over specific operation and faulty conditions and during a given period of time. This is rarely possible, because the relationship between dielectric strength and the many parameters entering in the time and conditions of the operation cannot be condensed in a single test.

It is common practice to test electrical components by applying a test voltage equal to the double of the operational voltage plus one kV: for example, a magnet operating at 500 V to ground may be tested at a voltage of 2 kV to ground. However, this general guideline has to be adjusted to the specific application and to the operational and faulty types of possible electrical stresses.

Magnets operating in accelerators are submitted, in addition to electrical stresses, to many environmental and mechanical stresses, which can be tolerated in case the magnet meets at least two conditions:

- it is correctly designed, with the appropriate materials, technologies and geometries;
- it is correctly manufactured.

The electrical tests intended to qualify the design and possibly the manufacture procedures are called *type tests* and, in many cases, they can be destructive. An example of type test are the well-known *crash tests* for cars, which aim at demonstrating that a specific design manufactured according to a specific process with given materials meets certain safety requirements.

Here we will focus on *acceptance tests for magnets*, intended to qualify individual objects, with two examples: the high-voltage test of a conductor to ground and the partial discharge measurements.

## 6.1 High-voltage test of a conductor to ground

Let us consider a magnet operated at 200 V in dc. We also assume its ground insulation scheme is designed with a minimum insulation thickness of 2 mm. By applying the rule of testing at two times the maximum voltage plus one kV, we would obtain a test voltage of 1.4 kV, which would seem high with respect to the applied voltage.

*Can we state that this test voltage is appropriate to ensure that the insulation to ground is well made?*

With 1.4 kV to ground the test would pass even if the insulation is cracked: the air in the crack, if dry, would need at least 2 kV and in ideal conditions even more than 3 kV (the dielectric rigidity of dry air, for 1 mm) to fail.

But, who cares if there is such a defect: in any case the magnet can withstand the required voltage!

Unfortunately, real life will care:

- the crack may dramatically propagate during time;
- a wet film may cover the crack surface, due to a water leak or just for ambient conditions.

In both cases the magnet would easily fail.

I leave it to the reader to state whether this test voltage is appropriate or not.

## 6.2 Partial discharge measurements

Partial discharge may be representative of defects or degradation of the insulation system.

PD can be studied with respect to:

- voltage level (to ground) triggering a flow of charges (PD inception voltage)
- voltage level stopping a flow of charges (PD extinction voltage)
- amount of charge involved in a single discharge
- energy dissipated in a single discharge
- time needed to produce inception or discharge at a given voltage
- characteristics of charge flow (in particular in ac)

In practice, for complex systems like magnet insulation, irrelevant defects may give signals not relevant to a correct diagnosis. On the other hand, certain defects, for example the ones produced by mechanical failure, may remain silent until shortly before failure.

PD techniques are thus important for large numbers (for example transformer windings), because one can count on statistical basis and exclude the components showing abnormal behaviour, or for repetitive and relatively simple geometries such as high-voltage cables.

## 7 Conclusions

Failure of the dielectric insulation of an electrical machine is often not dominated by electrical stresses but, in most cases,

- by a mechanical failure
- by an unsafe design with respect to environment or operation

A good designer ensures:

- the design correctly considers operation and fault conditions
- the test conditions are capable of identifying a defective manufacture.

**Appendix A: Superconducting magnets**

*A.1 Introduction*
As any electrical device, superconducting accelerator magnets require that active parts be dielectrically insulated from each other and from ground. Though the conductor resistance is null in the superconducting status, during transients, i.e, during magnet energization or during a quench, the voltage difference between adjacent cable turns and between coil and ground can rise to levels approaching or even exceeding the kV range. The large energy stored in superconducting magnets makes the event of an electrical short circuit a potential drama not only for the safety of the magnet itself but in certain cases for the whole accelerator. For example, the magnetic energy stored in a LHC dipole operating at 8.3 T is of about 7 MJ, corresponding to 1.5 kg of TNT (trinitrotoluene) [1]. Just a small a fraction of this energy is capable of melting the way from the superconducting cable to an austenitic steel vacuum chamber. In the case of operation in a helium bath, providing a good dielectric insulation may appear easier than in air, thanks to the high breakdown voltage of liquid helium, one order of magnitude larger than that of dry air, i.e, about 30 kV/mm [2]. However, in presence of vapour, as for example following the conductor heating during a quench or for systems not fully immersed in liquid helium, the breakdown voltage of helium is about one tenth that of air at the same pressure, and at sub-atmospheric pressures it decreases till its Paschen minimum of about 150 V [3]. Furthermore the insulating materials have to keep their dielectric and mechanical properties at cryogenic temperatures and under typically large mechanical and radioactive stresses.

*A.2 Dielectric strength*
In principle the dielectric strength required for a superconducting cable insulation, typically of the order of a few hundred volts between adjacent coil turns, may appear very comfortable with respect to the breakdown voltage of the insulating materials adopted. The reference breakdown voltage of a 25-micron thick polyimide tape is 7800 V and that of fully reacted pre-preg tapes is about 5 times smaller, strongly dependent on the material and on the degree of reaction. The challenges to design and implement a dielectrically reliable cable insulation are:
- in case of non-impregnated coils using cables insulated by tapes, ensure that the topology of the insulating scheme leaves at any place a sufficiently long path along the insulating tapes from one cable turn to the adjacent one. As a guideline, to provide a 1 kV dielectric strength at the beginning of a quench in gaseous helium at 1 bar the required surface length is of the order of 5 mm;
- in case of impregnated coils, ensure before the impregnation that each turn is correctly spaced from the other. In certain cases an electrical discharge test may be indicated to identify a defective assembly;
- in all cases ensure that the insulation will not break during magnet assembly and operation. In most cases, the main issue of preserving the dielectric insulation characteristics consists in preserving its mechanical integrity. As an example, for a given total thickness of the insulation, it may be preferable to wrap in contact with the cable surface the most robust tape, which is less sensitive to damage at cable edges so that the next tapes will lay on a smoother surface. In case of use of pre-preg tapes, or other composites, perform a thorough validation of mechanical characteristics and ageing at cold to evaluate possible degradation during the magnet lifetime. For example, using resin impregnated fibreglass tapes with too low percentages of resin may lead to a quick degradation of the composite after only a few cycles of operation.

Finally, it is important to remark that, in particular in magnets using multilayer coils, the voltage difference between adjacent layers can be much higher than the typical voltage difference between adjacent turns. In these cases it may be preferable to add additional interlayer insulators.

*A.3 Heat removal*

Another important property of a superconducting cable insulation is its thermal conductivity. The lower the thermal conductivity, the higher the heat which can be removed from the cable to the coolant for a given temperature difference. This limitation of course does not affect internally cooled cables, in which the superconductor is directly in contact with the coolant.

At cryogenic temperatures the study of heat transfer through cable insulation is rather complex, in particular because it depends on temperature, geometry, the status of helium and, depending on the cooling regime, on the status of the material surface. Furthermore, at the boundary between a solid material, such as cable insulation, and liquid helium the temperature is not continuous but there is a temperature difference at the interface, described by a thermal boundary resistance by Kapitza in 1941 [4]. An excellent review on Kapitza resistance can be found in Ref. [5].

Owing to these complexities, summarized in a review by Baudouy et al. [6], there are only a few measurements of the thermal conductivity of polyimide at low temperatures, the one mostly used as reference being that of Lawrence et al. [7]. In case of epoxy fibreglass composites data are even more difficult to interpret because of the variety of the compounds and of the surface status. A good reference can be considered the work of L. Imbasciati et al. [8].

In superfluid helium a powerful mechanism of heat transport can be obtained by providing micro-channels through the cable insulation: this technique has been adopted for the main superconducting magnets for the LHC and allowed to increase by about a factor two the power which can be removed from the cable to helium compared to a similar insulation without microchannels. Recently a new cable insulation scheme has been developed for a possible use for the new inner triplets for the LHC luminosity upgrade, providing a further increase of the heat transfer of about a factor of 5 [9,10].

**References for Appendix A**